\documentclass{aa}
\usepackage{graphicx}
\usepackage{natbib}
\usepackage{journals}
\usepackage{psfig}
\begin{document}

   \title{The European Photon Imaging Camera on XMM-Newton: \\
The MOS Cameras}

   \author{M.~J.~L.~Turner \inst{1}
	\and A.~Abbey \inst{1} 
	\and M.~Arnaud \inst{2}
	\and M.~Balasini \inst{13}  
	\and M.~Barbera \inst{10}
	\and E.~Belsole \inst{2}
	\and P.~J.~Bennie \inst{1}
	\and J.~P.~Bernard \inst{12}
	\and G.~F.~Bignami \inst{7}
	\and M.~Boer	\inst{3}
	\and U.~Briel \inst{4}
	\and I.~Butler \inst{5}
	\and C.~Cara \inst{2}
	\and C.~Chabaud \inst{3}
        \and R.~Cole \inst{1}
	\and A.~Collura \inst{10}
	\and M.~Conte \inst{7}
	\and A.~Cros \inst{3}
	\and M.~Denby \inst{1}
	\and P.~Dhez \inst{6}
	\and G.~Di Coco \inst{11}
	\and J.~Dowson \inst{1}
	\and P.~Ferrando \inst{2}
	\and S.~Ghizzardi \inst{7}
	\and F.~Gianotti \inst{11}
	\and C.~V.~Goodall \inst{5}
	\and L.~Gretton \inst{1}
	\and R.~G.~Griffiths \inst{1}
	\and O.~Hainaut \inst{12}
	\and J.~F.~Hochedez \inst{12}
	\and A.~D.~Holland \inst{1}
	\and E.~Jourdain \inst{12}
	\and E.~Kendziorra \inst{8}
	\and A.~Lagostina \inst{7}
	\and R.~Laine \inst{9}
	\and N.~La Palombara \inst{7}
	\and M.~Lortholary \inst{2}
	\and D.~Lumb \inst{14}
	\and P.~Marty \inst{12}
	\and S.~Molendi \inst{7}
	\and C.~Pigot \inst{2}
	\and E.~Poindron \inst{2}
	\and K.~A.~Pounds \inst{1}
	\and J.~N.~Reeves \inst{1}
	\and C.~Reppin \inst{4}
	\and R.~Rothenflug \inst{2}
	\and P.~Salvetat \inst{12}
	\and J.~L.~Sauvageot \inst{2}
	\and D.~Schmitt \inst{2}
	\and S.~Sembay \inst{1}
	\and A.~D.~T.~Short \inst{1}
	\and J.~Spragg \inst{1}
	\and J.~Stephen \inst{11}
	\and L.~Str\"{u}der \inst{4}
	\and A.~Tiengo \inst{7}
	\and M.~Trifoglio \inst{11}
	\and J.~Tr\"{u}mper \inst{4}
	\and S.~Vercellone \inst{7}
	\and L.~Vigroux \inst{2}
	\and G.~Villa \inst{7}
	\and M.~J.~Ward \inst{1}
	\and S.~Whitehead \inst{1}
	\and E.~Zonca \inst{2}
	}

   \offprints{M. J. L. Turner}
  
   \institute{Dept. of Physics \& Astronomy, Leicester University, LE1 7RH, UK.
	\and CEA/DSM/DAPNIA Service d'Astrophysique, CEA/Saclay, 91191 Gif-sur-Yvette Cedex, France.
	\and Centre d'Etude Spatiale des Rayonnements, 9 avenue du colonel Roche, BP 4346, 31028 Toulouse Cedex 4, France.
	\and MPE  D-85740 Garching, 8046, Germany.
	\and School of Physics and Astronomy, University of Birmingham, B15 2TT, UK.
	\and Laboratoire pour l'Utilisation du Rayonnement Electromagnetique, Bat 209 D, Universite Paris Sud, 91405 Orsay, France.
	\and IFC Milan, 20133 Milano, Italy.
	\and IAAP Tuebingen, D-72076, Germany
	\and PX ESTEC, Postbus 299, 2200 AG Noordwijk, Holland.
	\and Osservatorio Astronomico di Palermo, Palermo, 90134, Italy.
	\and ITESRE, 41010 Bologna,Italy.
	\and Institut d'Astrophysique Spatiale, Bat 121, Universit\'{e} Paris Sud, 91405 Orsay, France
	\and Laben S.p.A, S.S. Padana Superiore, 290, 20090 Vimodrone, Milano, Italy
	\and Space Science Department, ESTEC, 2200 AG Noordwijk, Holland
}

   \date{Received; accepted }

   \authorrunning{M. J. L. Turner et al.}

\abstract{
The EPIC focal plane imaging spectrometers on XMM-Newton use CCDs to record the
images and spectra of celestial X-ray sources focused by the three X-ray 
mirrors. There is one camera at the focus of each mirror; two of
the cameras contain seven MOS CCDs, while the third uses twelve PN CCDs, 
defining a circular field of view of 30$^{\prime}$ diameter in each case.
The CCDs were specially developed for EPIC, and combine high quality
imaging with spectral resolution close to the Fano limit. A filter
wheel carrying three kinds of X-ray transparent light blocking filter, a fully
closed, and a fully open position, is fitted to each EPIC instrument. The CCDs
are cooled passively and are under full closed loop thermal control. 
A radio-active source is fitted for internal calibration. Data are 
processed on-board to save telemetry by removing cosmic ray tracks, and 
generating X-ray event
files; a variety of different instrument modes are available to increase the
dynamic range of the instrument and to enable fast timing. The instruments
were calibrated using laboratory X-ray beams, and synchrotron
generated monochromatic X-ray beams before
launch; in-orbit calibration makes use of a variety of celestial 
X-ray targets. The
current calibration is better than 10\% over the entire energy range of 0.2 
to 10 keV.
All three instruments survived launch and are performing nominally in orbit. In
particular full field-of-view coverage is available, all electronic modes work,
and the energy resolution is close to pre-launch values. Radiation
damage is well within pre-launch predictions and does not yet impact on the 
energy resolution. The scientific results from EPIC amply fulfil pre-launch 
expectations.
}

   \maketitle

%

\section{Introduction}

\begin{figure}
\centering
\includegraphics[width=7cm]{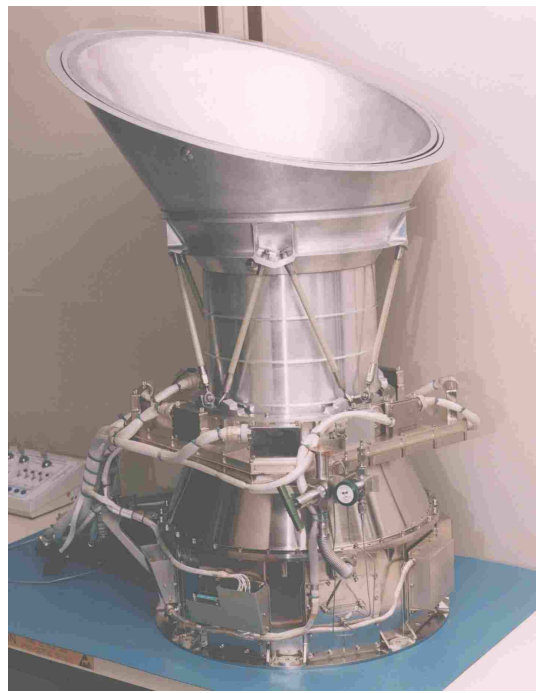}
\caption{The long conical radiators of the MOS enable the radiating 
surface to reach the plane of the spacecraft thermal shield to avoid 
parasitic heat-loads. }
\label{mos}
\end{figure}

The EPIC instrument on XMM-Newton \citep{jansen01} provides focal plane
imaging and spectrometry for the three X-ray telescopes. Each telescope has an 
objective comprising a nested, 58 shell, Wolter 1 X-ray mirror \citep{aschenbach01}, of focal length
7.5 metres, and geometric effective area 1500 cm$^{2}$; there is one EPIC
at the focus of each telescope. Two of the telescopes are fitted with the X-ray
gratings of the Reflection Grating Spectrometer \citep{denherder01}. 
The gratings divert
50\% of the flux out of the EPIC beams; with allowance for structural
obscuration, 44\% of the original flux reaches two of the EPIC cameras; these contain 
MOS CCDs \citep{short98} and are referred to as the MOS cameras.
The third telescope has an unobstructed beam; the EPIC instrument at the focus
of this telescope uses PN CCDs \citep{struder01} and is referred to 
as the PN camera. 
All three cameras 
have an
identical forward section that contains a filter wheel, door, 
calibration source,
radiation shielding, the interface to the spacecraft focal plane 
bulkhead, and the
internal bulkhead that forms part of the camera vacuum enclosure. The 
rear part
of each camera that contains the CCDs and the cooling system is 
different in
construction for the MOS (figure~\ref{mos}) and PN cameras. EPIC also 
includes the EPIC Radiation Monitor System, to record the ambient proton 
and electron flux \citep{boer96}. It provides warning of a radiation 
flux increase to provide for automatic shut down of the instrument.
This paper describes the common items and the
MOS cameras, while an accompanying paper, by \cite{struder01}
describes the PN camera.

\section{The EPIC Camera Configuration}

Each EPIC camera has three sections: the stand-off structure that contains the
filter wheel, door, calibration source, interface to the spacecraft, 
the radiation
shield and the internal vacuum bulkhead; the cryostat section that contains the
CCDs and electronic interfaces, and the radiator that provides the cooling for
the CCDs. The stand-off-structure is identical for all three cameras, and is
described here; the remaining sections are different for PN 
\citep{struder01} and MOS (see below).

\subsection{The Stand-off Structure}

\begin{figure}
\centering
\includegraphics[width=8cm]{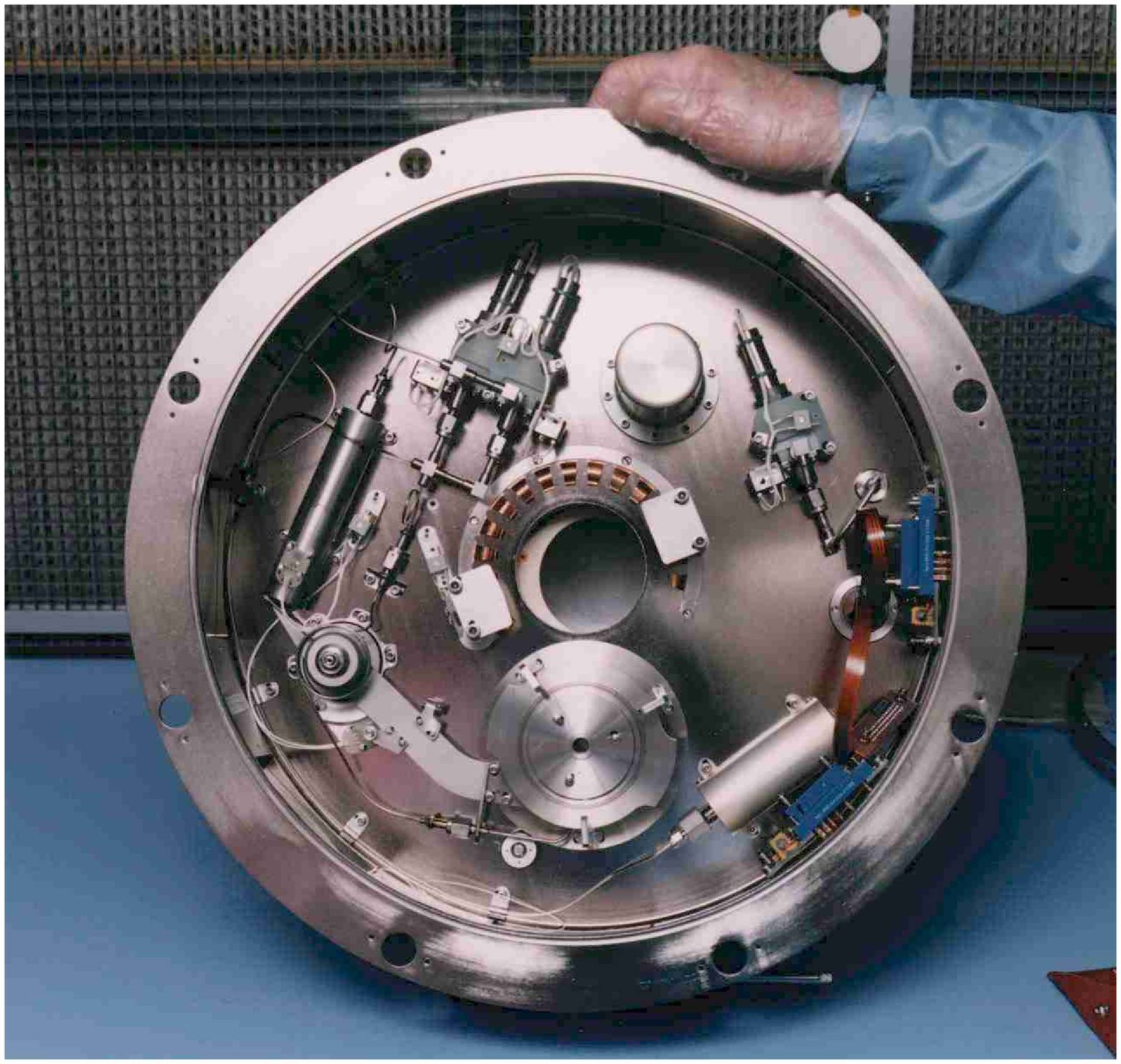}
\includegraphics[width=8cm]{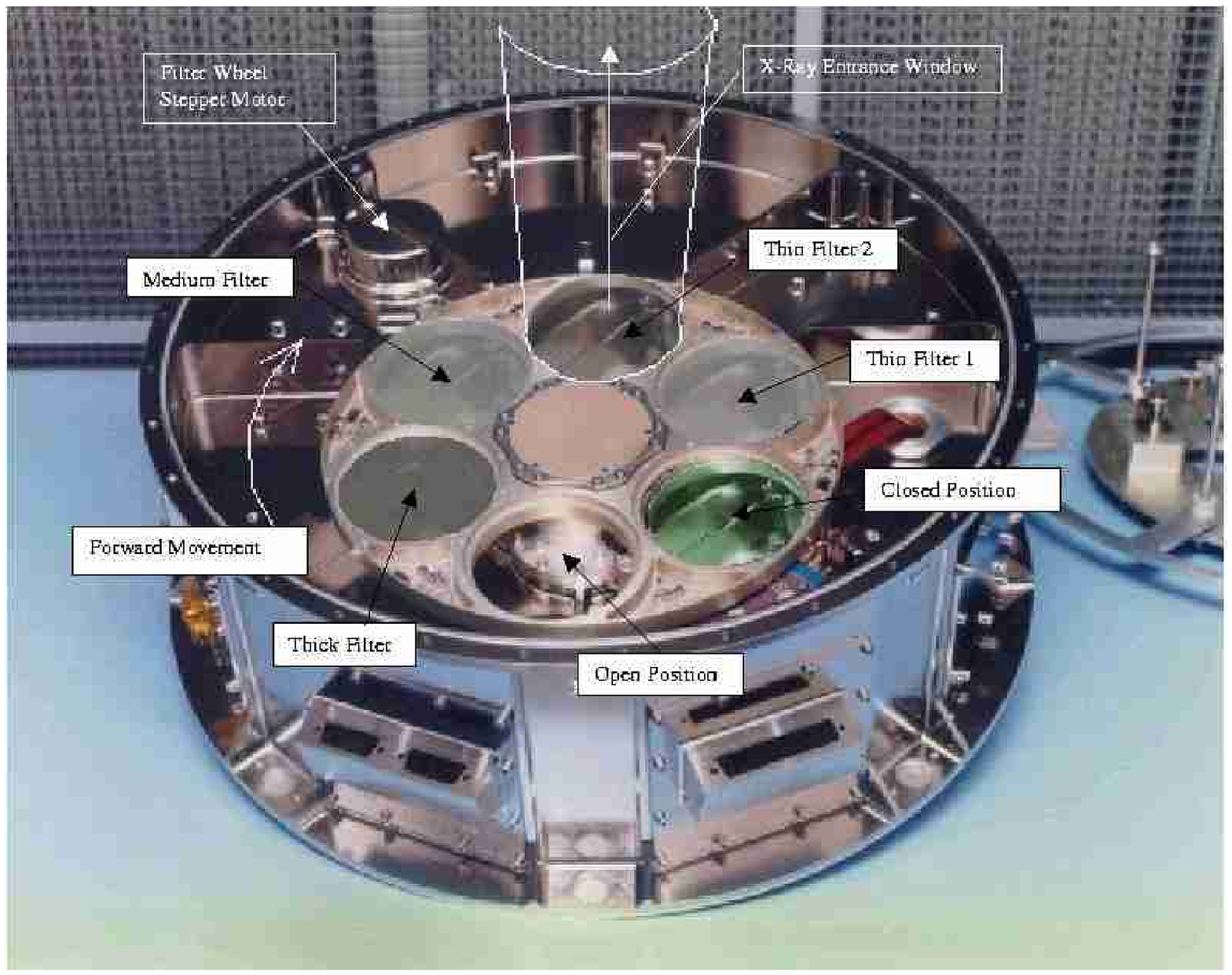}
\caption{The EPIC Stand-off Structure. The aluminium machining is nickel
plated to reduce vacuum out-gassing. It houses the filter wheel, calibration
source and door; the bulkhead is part of the vacuum enclosure. It is located on
the spacecraft bulkhead using drilled bushes to bring the optical centre of the
camera to the optical axis of the mirror.}
\label{sos}
\end{figure}

The structure is shown in figure~\ref{sos}.
The filter wheel with its gears and motor is located on the vacuum side of the
bulkhead (figure~\ref{sos}, bottom panel). It has six 
locations for
filters, and six small apertures through which the calibration source can shine
onto the CCDs. The source itself is mounted in the bulkhead, on the
vacuum side. The aperture in the bulkhead through which the X-rays reach the
CCDs is closed before and during launch by the door. The door is
mounted on a spring-loaded arm, and is clamped by an integral set of double
stainless steel bellows that, when pressurised, press a plate on to an O-ring
mounted in the bulkhead. When the nitrogen pressure of 4 bar is released, the
bellows retract, and the door swings out of the aperture, under action of the
spring. The CCDs are protected from damaging proton irradiation by
at least 3 cm of aluminium shielding in all directions, except the 
field of view.

\subsection{The MOS Cryostat}

The cryostat interfaces to the stand-off
structure. The cold-finger is supported by aluminium and GFRP components
forming a doubly insulated structure with a secondary thermal shield, thermally
linked to the outer radiator. The CCDs are mounted on the end of the cold
finger, and the rear 3 cm shielding is integral with the CCD mounting plate. The
CCDs themselves are mounted on individual packages. Each package slides in a 
channel in the
mounting plate; the CCDs are located accurately by this channel 
(figure~\ref{focalplane}). The
heaters for thermal control are incorporated in the mounting plate, which is at
the operational temperature of the CCDs ($\sim$170 K). Electrical 
connections to the
CCDs and to the heaters are made with thin flexible printed circuits, to
provide a thermal break. The preamplifiers and other electrical components are
mounted on an external conventional multi-layer PCB. The PCB also acts as the
vacuum feed-through to bring signals and power into the cryostat. The 
cryostat is closed at the rear with two 50 micron stainless steel foils 
that link the outer and inner thermal shields to the cold finger; 
they provide a vacuum and EMC tight connection with minimal thermal
conductivity.

\subsection{The MOS Radiators}
The MOS cryostat is cooled passively \citep{butler00} by the system of three 
nested conical radiators. This shape is necessitated by the 
distance between the CCD focal plane and the top of the spacecraft sun-shield. 
The outer radiators provide successive thermal shielding for the inner 
radiator which is attached to the CCD via the cold finger. This inner 
radiator is mechanically 
supported by the cold finger and has no contact with the other radiators.
The secondary radiator is connected 
to the thermal shroud surrounding the CCD and cold finger via a flexible thermal
 coupling.
Doubly aluminised Mylar foils independently suspended between successive 
radiator cones provide additional shielding thus enhancing performance and 
robustness against deterioration in space of the thermal finish of surfaces.
The radiators are made of spun aluminium of 0.6 mm thickness.  The inner radiator
 is painted with white epoxy paint to give maximum emissivity, coupled with low 
absorptivity for Earth albedo during perigee.
The cooling system provides a minimum temperature at the CCD of 140 K.


\section{The Electronics}

\begin{figure}
\includegraphics[width=9cm]{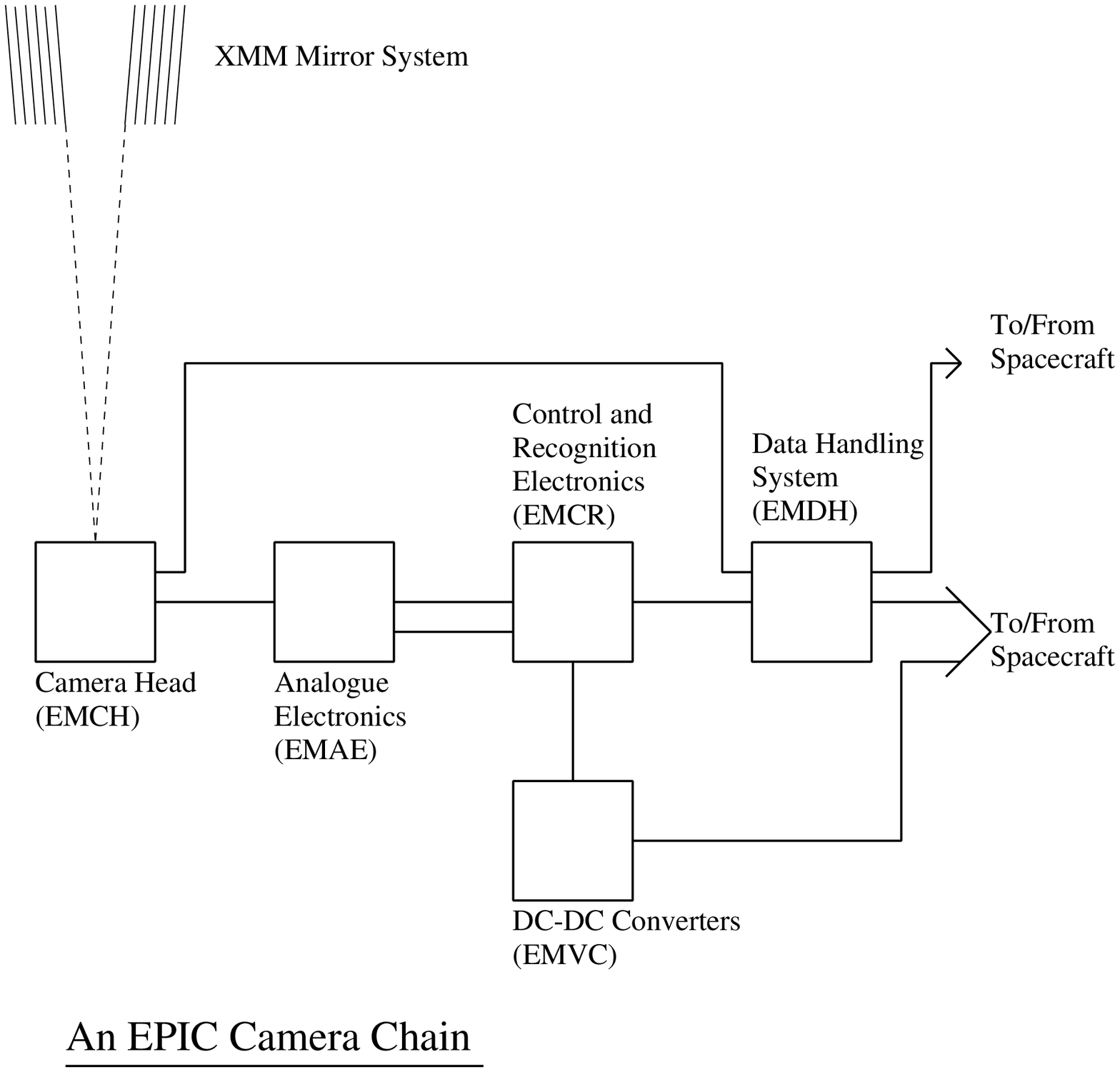}
\caption{The EPIC MOS electronics block diagram}
\label{emcs}
\end{figure}

The electronic block diagram is shown in figure~\ref{emcs} \citep{villa96}. 
The preamplifiers are 
mounted on the cryostat PCB, integral with the camera. These interface via 
the harness to the analogue electronics box EMAE. This conditions the 
signals and contains the CCD clock sequencers. The system is partially 
redundant with one set of eight channels reading out the seven CCDs in a 
camera. There are two readout nodes on each CCD. The outer six CCDs are 
normally read from one node, and the central CCD can be read from one node, or 
for faster response by two in parallel. The second of each pair of 
processing chains is connected to the 
alternate nodes of the outer CCDs; multiplexers allow crossover in the case of 
failure.  The EMAE also contains the clocking sequencers that generate the 
trains of pulses to transfer the charge released by the X-rays to the readout 
node.  The control and recognition unit EMCR \citep{pigot00,ferrando99} has the 
main function of converting the image read out from each CCD into an X-ray 
photon event list. This uses fast logic (an ASIC) to subtract offsets 
so that the energy scale starts at zero; it identifies geometric 
patterns of pixels using a look up table, programmable from the ground, to 
distinguish between those produced by X-rays and those produced by cosmic rays.
It also adds up the charge over the pattern in each event to enable the X-ray 
energy to be determined accurately. The output is an event list with up to four
different energy parameters attached, together with the pattern shape.  
Converting the image into an 
event list results in significant data compression, because of the sparse 
nature of the X-ray image. The pattern recognition enables cosmic rays to be 
rejected on-board. The other function of the
EMCR is to store and transmit CCD clock sequences for the different modes to
the CCD clock sequencers in the EMAE; it also provides the control signals to
the filter wheel and timestamps CCD frames.
Power conditioning and distribution is provided by the voltage controller, 
EMVC. The data handling and interface to the spacecraft data system and 
telemetry is
provided by the EMDH \citep{villa96}. This receives data from the EMCR and
conditions it for the telemetry. It receives, distributes, and 
executes commands;
it also contains the on-board software that controls the instrument modes. The
EMDH controls all instrument safety aspects and is the interface to the
spacecraft for software upload, command and data handling.

\section{The CCDs}

\begin{figure}
\centering
\includegraphics[width=6cm]{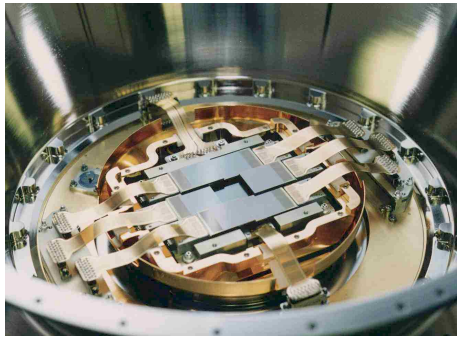}
\caption{The CCDs with their flexible PCB links shown mounted in the
cryostat}
\label{focalplane}
\end{figure}

There are seven EEV type 22 CCDs in the focal plane of each MOS camera 
\citep{short98}, arranged as shown in figure~\ref{focalplane}. The central 
CCD is at the 
focal point on the 
optical axis of the
telescope while the outer six are stepped towards the mirror by 4.5 mm to 
follow approximately the focal plane curvature, and improve the
focus for off-axis sources. The CCDs are buttable with a dead region of less 
than 300 microns
wide on three sides; to minimise the dead space, adjacent CCDs are stepped by
about 1mm to overlap by 300 microns.
The EEV CCD22 is a three-phase frame transfer device on high
resistivity epitaxial silicon with an open-electrode structure; it 
has a useful quantum
efficiency in the energy range 0.2 to 10 keV. The imaging area is 
$\sim$ 2.5 x 2.5 cm, so that a mosaic of seven covers the focal plane 
62 mm in diameter, equivalent to 28$^{\prime}$.4. The imaging
section has 600 x 600, 40 micron square, pixels; one pixel
covers 1.1 x 1.1 arc seconds on the field of view; 15 
pixels cover the mirror PSF half energy width of $15''$. The readout
register is split into two sections, ending in a readout
node. The full CCD image can be read out using either node, or
read out using both nodes simultaneously, to halve the
readout time.
The low energy response of the conventional front illuminated CCD is
poor below $\sim$700 eV because of absorption in the electrode
structure. For EPIC, one of the three electrodes has been enlarged to 
occupy a greater
fraction of each pixel, and holes have been etched through this enlarged
electrode to the gate oxide. This gives an 'open' fraction of the 
total pixel area of
~40\%; this region has a high transmission for very soft X-rays that would
otherwise be absorbed in the electrodes. In the etched areas, the surface
potential is pinned to the substrate potential by means of a 'pinning 
implant'. High energy 
efficiency is
defined by the resistivity of the epitaxial silicon ( around 4000 Ohm-cm). The
epitaxial layer is 80 microns thick (p-type). The actual mean depletion of 
the flight CCDs is between 35 and 40 microns:
the open phase region of each pixel is not fully depleted. The two MOS 
cameras are 
arranged on the spacecraft focal
plane bulkhead so that the CCDs are orthogonal. This means that the 
300 micron gaps
between the outer CCDs in one camera are covered by their opposite 
numbers in the other
camera. The gaps around the central CCD in either camera exactly overlap:
so a source falling in this gap in MOS1 will also be in the 
gap in MOS2. The PN CCDs are arranged at 45 degrees to the two MOS 
cameras so the field
of view is fully covered among the three EPIC cameras.


\section{The EPIC MOS Readout Modes}

The basic readout speed of the MOS CCDs is 2.6 seconds; this is the same as the
integration time, and is available continuously on all seven CCDs in each 
camera. The only reason to change this is if the source is bright: pile up may 
be a problem; or there may be an opportunity for fast timing observations. 
The modes comprise 'large and small window mode', 
'refreshed frame store mode' \citep[see][]{willingale01} and 'timing mode'. 
Window 
mode can be applied to the central CCD and independently to the peripheral CCDs
in pairs (not used at present). A window is defined on the
CCD (windows need not be central on the CCD) and rows and pixels 
outside this window are discarded on readout. Since
most of the readout time is taken up by measuring the pixel charge
accurately, this results in a much faster readout and integration. 
Two window sizes are currently implemented:
Large window 300 $\times$ 300 pixels centred - 0.9 s integration time.
Small window 100 $\times$ 100 centred - 0.3 s integration time.
In full frame mode, an electronic chopper can be implemented - "Refreshed frame
store mode" which discards most of the X-rays collected during 
readout and integrates 
for 0.2 seconds. The overall cycle time is 2.9 s leading to a duty cycle of 
1:14.5.
Selection of the window size is a compromise between readout speed and field of
view coverage: source extent, and the PSF wings may require a larger window. 
The window is only applied to the central CCD, the peripheral CCDs remain fully
operational and the only area lost is the region outside the window on the 
central CCD.
For timing mode - also known as "Fast mode" - 100 rows in a central window 100 
pixels wide are compressed into one dimension, parallel to the CCD readout 
register to give fast time slices of the incoming X-ray flux. The CCD 
output node 
is only reset once per row to give a constant timebase (unlike the standard 
"reset-on-demand"), and the EDU event processing is simplified so less data has
to be telemetered. MOS1 and MOS2 are orthogonal, so two projections at right 
angles are available if both cameras are in timing mode. This mode gives a 
timing resolution of 1.75 ms per row or time slice. An experimental full width 
timing mode can also be implemented to give around 10 ms resolution.

The choice of mode is based on the extent of the source and its 
brightness; for the majority of sources the basic mode is acceptable, 
and window or timing modes are selected to deal with pile-up in 
bright sources \citep{ballet99}.
For source strengths below a few counts per second, the flux within the PSF 
is small enough that the probability of interference between the 
pixel patterns of individual photons is small. In this case there is 
no error in either flux or photon energy determination, and pile-up 
is negligible. As the flux increases the patterns from successive 
photons detected in the 2.6 s integration time start to interfere, and 
pile-up becomes significant. In the MOS cameras the dominant effect 
of pile up is a loss of flux; there is little distortion of the 
energy spectrum. This is because early pile-up results in the 
generation of non-X-ray-like event patterns, caused by contact 
between two genuine patterns; such events are removed automatically 
from the data stream by the EMCR, and do not appear in the spectrum. 
The PSF has a very sharp core, and this means that the 
flux-loss associated with pile-up occurs first in the core of the 
PSF, while the wings remain unaffected: a typical indicator of 
pile-up is a source cross section in the image with a depressed 
central region. Since flux removal is the main effect it is 
moderately safe to form the spectrum from the remaining flux even 
though the centre of the PSF is missing. In general it is safe to err 
towards imaging mode or large windows, unless the source is very 
bright.

Timing is related to the CCD readout and through the electronics to 
the spacecraft clock. The precision of timing is limited to the 
effective integration time. The accuracy is achieved via the 
spacecraft clock to UTC, and for a given sample there is a $\pm 40~\mu$s 
jitter caused by the fine time code sampling of this 
clock.


\section{The EPIC MOS Response Function}

\begin{figure}
\includegraphics[width=8cm]{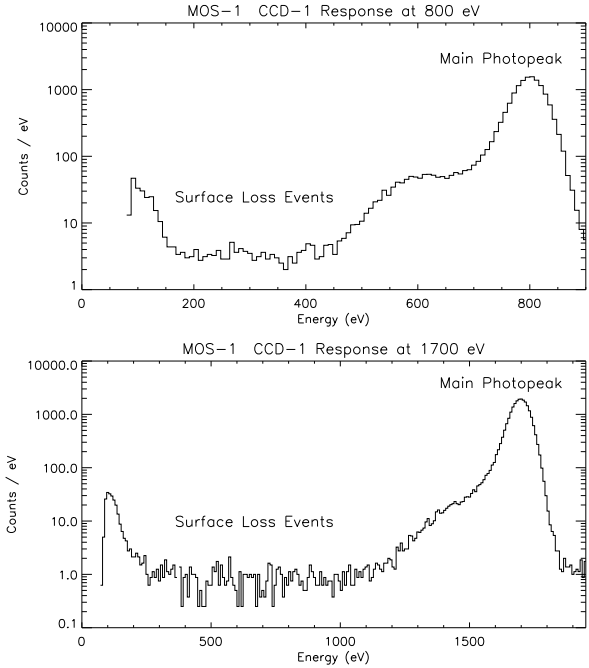}
\caption{The energy redistribution function of the MOS CCDs for
monochromatic X-rays of different energies, as measured at the Orsay
synchrotron}
\label{rmf}
\end{figure}

\begin{figure}
\rotatebox{-90}{\includegraphics[width=6cm]{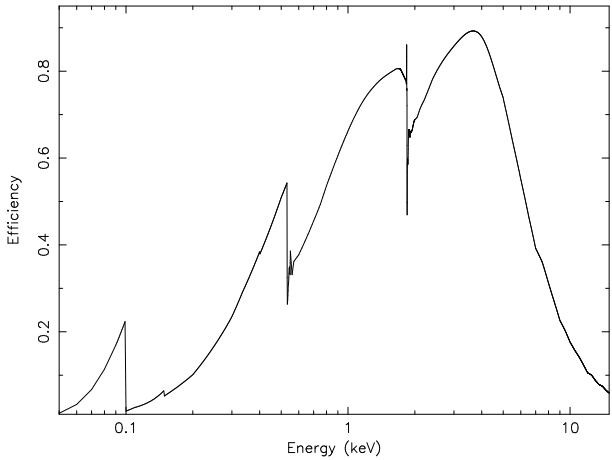}}
\caption{The X-ray quantum efficiency of the EPIC MOS CCDs based on the
Orsay synchrotron measurements and celestial source measurements}
\label{qe}
\end{figure}

\begin{figure}
\centering
\includegraphics[width=8cm]{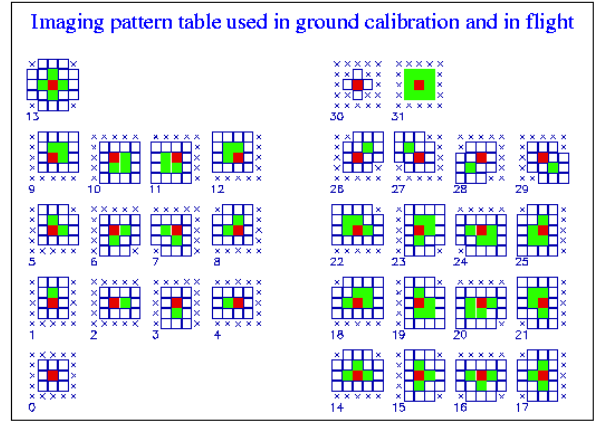}
\caption{The event patterns recognised by the EMCR in the imaging modes. The
coding is the following : the red pixel is the centre pixel, its signal is
above threshold and is the largest signal in the 3 $\times$ 3 inner matrix;
the green pixels have signals above threshold; the white pixels have signal
below threshold; the crosses indicate pixels not considered.}
\label{pattern}
\end{figure}

The CCDs image the entire field of view and give an
energy value and position to all X-ray photons detected. 
The flux and spectrum of X-ray sources in the field of view can be determined
from the charge measured in the individual photon event by the EMCR, the
redistribution matrix of the CCD, and its quantum efficiency. These have been
determined by a combination of ground calibration using the Orsay 
synchrotron \citep{pigot99,trifoglio98}, and celestial source calibration 
since launch. The current relative
accuracy of calibration is better than 10\% over the energy range 
from 0.2 to 10 keV.
X-ray photons falling on the CCD can be absorbed in the electrode structure, in
the depleted silicon, or in the field-free un-depleted silicon. A 
photon absorbed
in the electrode structure is lost; one converting in the depleted region is
detected with all its energy; one converting in the field free region 
is detected,
but some energy may be lost. These factors go together to make up the quantum
efficiency of the device. In the case of the CCD22 the open phase produces
additional complication because the low energy photons detected can interact
either in the pinning implant, or in the depleted silicon beneath it. 
In the implant
region a certain fixed amount of charge can be lost. So for soft photons there
may be a low energy 'shoulder' on the peak observed for a monochromatic X-
ray beam, or even a double peak; this degrades the energy resolution. 
These surface loss events have been carefully calibrated using data
from monochromatic X-ray beams in the Orsay synchrotron \citep{pigot99,dhez97},
 and celestial sources; some examples are shown in figure~\ref{rmf} for 
different energies. This effect is fully taken into account in the 
calibration files.
The quantum efficiency of the CCDs is a smooth function except near the edges
of silicon and oxygen; the carbon and aluminium edges  are apparent in
the thin and medium filter
responses, tin appears as well in the thick filter (the gold edges 
of the mirror are apparent in the overall quantum
efficiency). The response near these edges was measured using 
different beams at the Orsay
synchrotron. The measurements have been linked together using
celestial sources, especially BL Lac objects that have a smooth featureless
spectrum, and the Crab nebula spectrum, which is likewise featureless. The
current quantum efficiency curve is shown in figure~\ref{qe}. It 
varies a little from
CCD to CCD at very low energies.
The energy assigned to a photon is calculated from the charge in the pixel
pattern. Most of the photons detected in EPIC MOS give single pixel events.
For two, three, and four pixel events, the charge is summed over the relevant
pixels. The event patterns
recognised by the EMCR are shown in figure~\ref{pattern}. Patterns 
zero to twelve are X-ray events while the rest are not, with the 
exception of pattern 31. This is
formed by cosmic rays and also by deeply interacting energetic X-rays. 
All of these effects are included in the published calibrations of 
the EPIC MOS;
we are working to improve calibration, and this will be reflected in subsequent
issues.

\section{The Filters}

\begin{figure}
\centering
\rotatebox{-90}{\includegraphics[width=6cm]{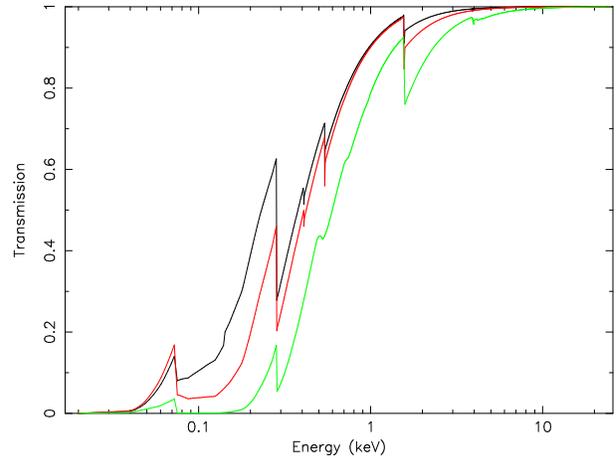}}
\caption{The X-ray transmission of the thin(black), medium(red) and 
thick(green) filter}
\label{xrayfilt}
\end{figure}

The EPIC cameras are provided with light
and UV blocking filters. There are four filters in each camera 
\citep{villa98,stephan96}. Two are thin filters made of 1600~\AA~
of poly-imide film with 400~\AA~of 
aluminium evaporated on to one side; one is the medium filter
made of the same material but with 800~\AA~of aluminium deposited on
it; and one is the thick filter. This is made of 3300~\AA~thick 
Polypropylene with 1100~\AA~of
aluminium and 450~\AA~of tin evaporated on the film. Poly-imide is
opaque to UV, but polypropylene is not, hence the tin. 
The filters are
self-supporting and 76 mm in diameter. The remaining two positions on the
filter wheel are occupied by the closed and open positions respectively. The
closed position has 1.05 mm of aluminium, and the open position has nothing.
The former is used to protect the CCDs from soft protons in orbit, while the
open position can be used for observations where the light flux is 
very low, and
no filter is needed.
Light falling on the CCDs
increases the baseline
charge level in the exposed pixels; this adds on to charge measured when an
X-ray photon falls on an affected pixel. It only has an effect where X-ray
photons fall, e.g. a bright optical
point source in
an image will have no effect on X-ray sources unless it coincides
or overlaps with one of them; scattering and the wings of the PSF have to be
taken into account in defining overlap. Diffuse illumination will have an
effect over the whole focal plane. In general diffuse optical 
illumination is very
low in EPIC, so the filter should be selected for the optical flux of 
the source being observed.
The X-ray transmission as a function of photon energy are shown in 
figure~\ref{xrayfilt}.

\section{On-Board Calibration Source}

\begin{figure}
\centering
\rotatebox{90}{\includegraphics[width=6.5cm]{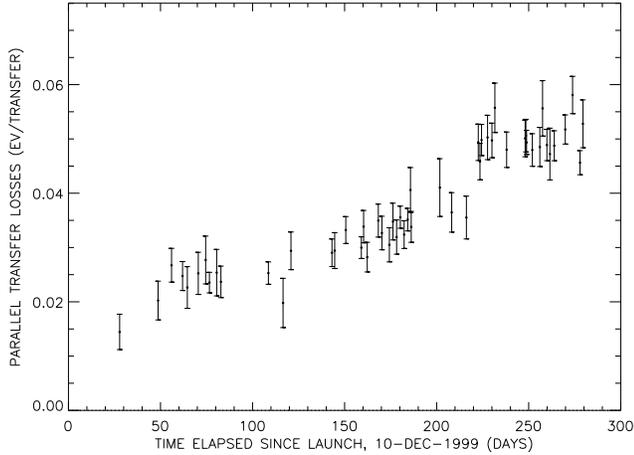}}
\caption{The evolution of spatial charge loss with time since launch in eV
channels per pixel as measured in the Mn K$_{\alpha}$ line. A step at day 
$\sim$ 220 can be seen corresponding to a very large solar flare.}
\label{cti}
\end{figure}

A calibration source is fitted to 
each camera to
allow the absolute energy to be calculated to an accuracy of a few eV; this is
important for line velocity measurements. The source uses iron 55, and provides
Al K$_{\alpha}$, Mn K$_{\alpha}$ and Mn K$_{\beta}$ at 1.5, 5.9 and 
6.5 keV respectively. It can illuminate the whole focal plane, for any 
filter position. This enables the gain of the system to be 
measured for the same conditions and time as the observation. 
It should be noted that for all CCDs there is a dependence of measured
charge on the position where the photon was detected within the CCD. This
is caused by charge loss during transfer of the photon event from its
detected position to the readout node.  This is expressed by the Charge
Transfer Efficiency of the CCD.  The monochromatic synchrotron data was
used to determine and correct the average charge transfer losses for the
MOS CCD readout, prior to generating the response function.  The CTE is
degraded by radiation damage in-orbit, and already this can be
detected after nine months in orbit, see figure~\ref{cti}. However, this 
degradation, if left uncorrected, would represent a 
loss of spectral resolution of less than 2\%.
The spatial correction employed will be progressively updated during the
life of the EPIC MOS cameras.  

\section{Radiation Damage and Precautions}

\begin{figure}
\centering
\rotatebox{90}{\includegraphics[width=6cm]{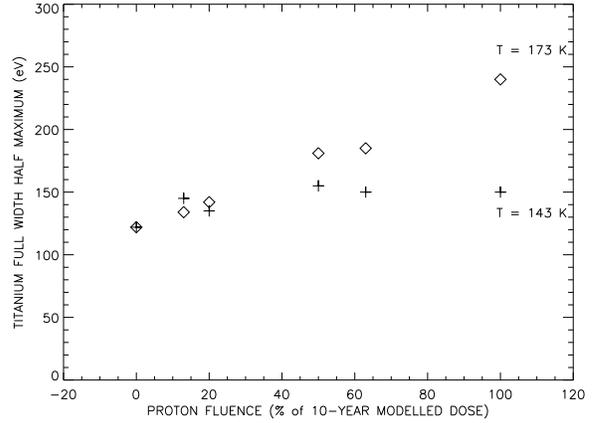}}
\caption{The recovery of radiation damage on CCD 22 with cooling. The damage 
was induced by 10 MeV protons at the Birmingham University Cyclotron. 
The full scale is approximately equivalent to ten years operation in
orbit, based on pre-launch calculations.}
\label{temp}
\end{figure}

\begin{figure}
\centering
\includegraphics[width=6cm]{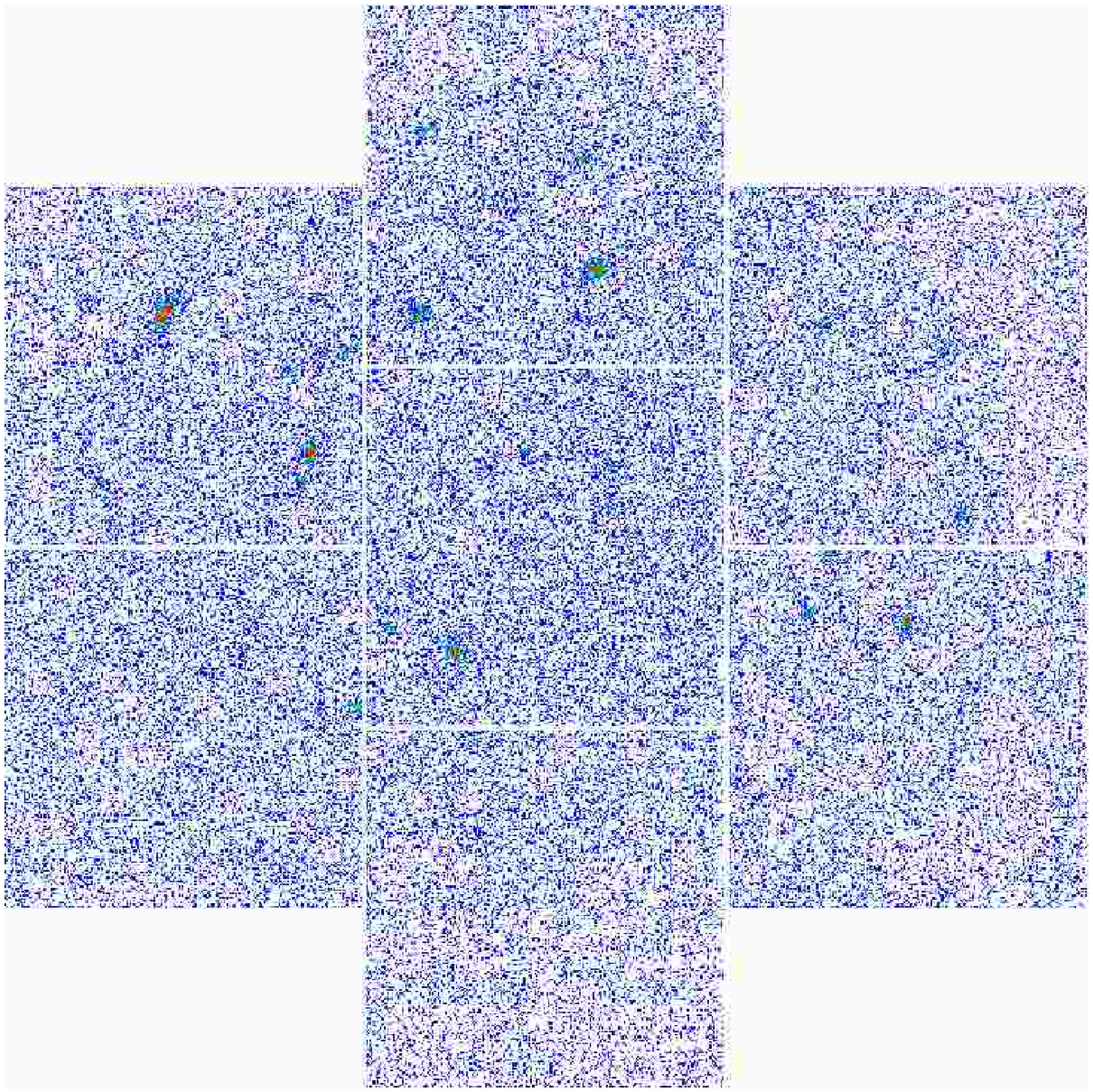}
\includegraphics[width=6cm]{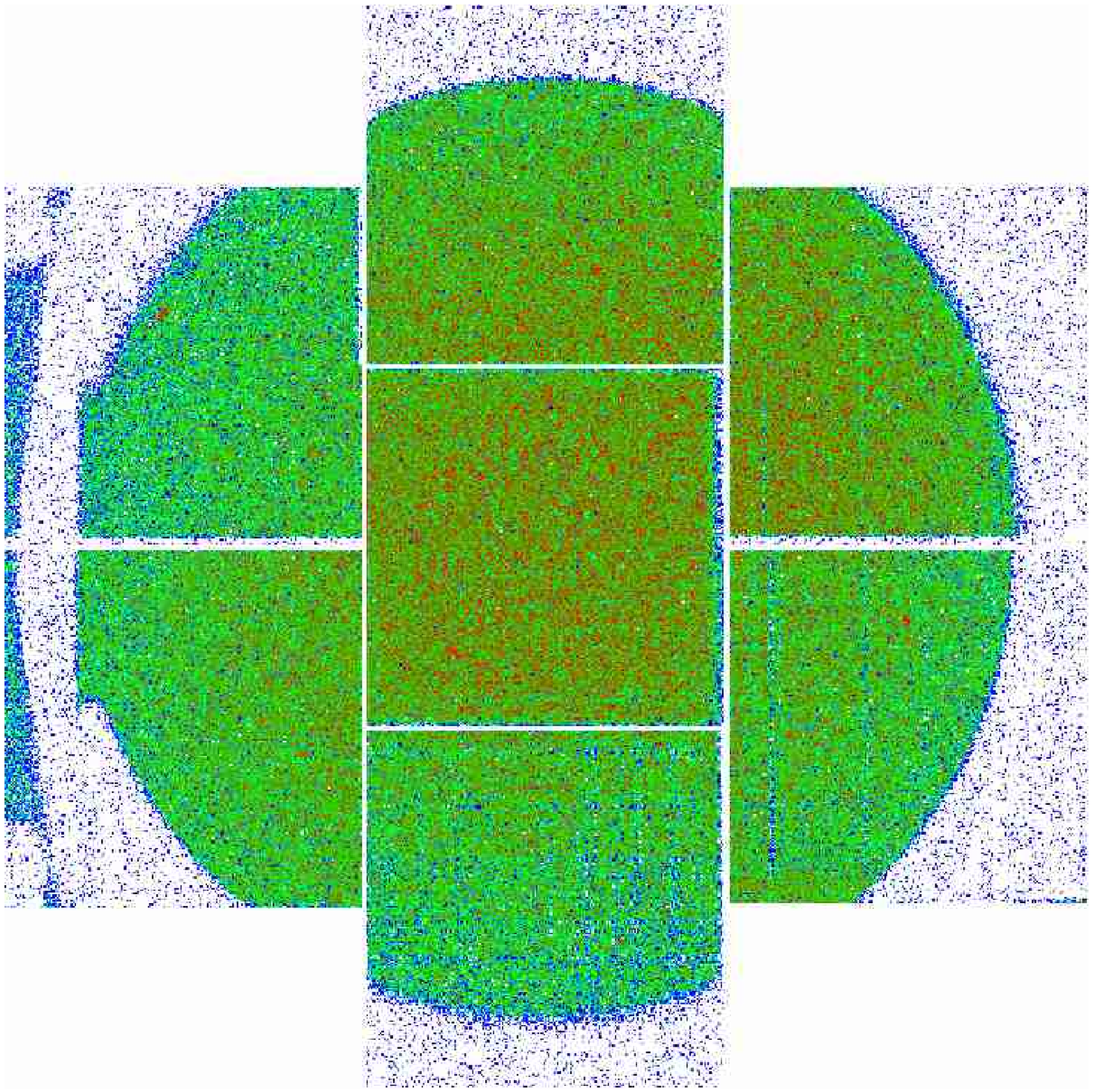}
\caption{Two images made with different soft proton backgrounds. The
brightness of the background where the flux is high (bottom) is sharply 
shadowed by the metalwork of the camera. In the low flux image (top) the 
background is not enough to delineate the metalwork.}
\label{protons}
\end{figure}

EPIC CCDs (MOS \& PN), have a high degree of immunity to ionisation 
damage, they are however sensitive to displacement-damage by 
protons with energy between 150 keV and 10 MeV: atoms in the silicon
lattice are displaced, creating a damage site of
locally low
potential wherein charge can be trapped. Damage sites in the buried channel,
through which the X-ray generated charge is transferred to the read-out node,
hold back some electrons; delayed electrons are not included when the charge
due to the X-ray event is determined. They are released later by 
thermal motion,
the trapping time depending inversely on the temperature of the CCD. By
increasing the density of charge passing through the buried channel, 
the loss can
be reduced for normal operating temperatures; this occurs because a leading
electron fills the trap, which then remains inactive, until the 
electron is released.
The other electrons pass the inactive trap safely. By confining the 
electrons to a
narrow buried channel during transfer, fewer traps are encountered and the
charge-loss is further reduced.  This is the reason for the current high CTE of
the MOS CCDs \citep{holland90}.
If the CCD is operated at a very low temperature \citep{holland92} 
then the trapping
time becomes very long, and the trap is essentially masked permanently. The
MOS cameras have reserve cooling power to allow operation at 140 K.
This precaution is not yet necessary, but will, when damage becomes severe,
recover most of the consequent CTE loss as shown in figure~\ref{temp}.
Presently the CCDs are operated at
170 K, to minimise condensation build-up on the CCDs.
Because the most damaging protons are non-relativistic (they have a high
displacement cross-section), the 3 cm shielding is effective (below 30 MeV). 
More energetic protons can penetrate the shielding, but their
flux and cross section are much lower. In combination with the extra cooling
this shielding was expected to be sufficient to protect the CCDs over the 
ten-year lifetime of XMM-Newton, taking into account the orbital evolution.
It is now known that very soft protons can be 
focused by the X-ray mirrors, and reach the CCDs, by-passing the shielding 
\citep{rasmussen99}.
These protons stop
in less than a micron of material; in MOS CCDs the
buried channel lies only 0.5 microns below the surface, so they can
cause very significant damage; two orders of magnitude greater than 
10 MeV protons. The
large effective area of the XMM-Newton mirrors makes this focused soft proton
flux very high, and the potential for damage very serious. 
Fortunately the closed
position of the filter wheel places 1.05 mm of aluminium in the focused beam
and completely stops the protons. For all perigee passages the
filter wheel is rotated to the closed position before the spacecraft 
encounters the
outer radiation belts. This has prevented damage by soft protons in 
the belts.
 Soft protons are also found outside the belts, probably accelerated 
in the magnetopause. When the flux of these is high the filter wheel is 
rotated to the closed position
to protect the
CCDs. Lower fluxes can contribute to the background in the image during
observations. They stop close to the surface of the silicon and generate
single pixel events with a flat energy spectrum; they are indistiguishable from
X-rays. It is not known if the prevalence of these clouds of soft protons will
change as solar activity decreases, or as the orbit of XMM-Newton evolves. On
average several hours per orbit have a high soft proton background, or are lost
completely because the filter wheel has to be closed. Figure~\ref{protons} 
shows an image
collected while the soft proton flux was relatively high compared with a low
proton flux image.
There have been two large solar flares so far since the launch of XMM-Newton.
These generate high fluxes of penetrating protons that cause CCD damage
through the shielding; they are not focused by the mirrors. After the 
first flare a
measurable degradation of the CTE was observed, equivalent to several
months' passages through the radiation belts. Thus the CTE falls 
steadily due to
the latter effect while solar flares generate steps in the curve, 
as shown in figure~\ref{cti}.
If the CTE is degraded seriously, and extra cooling cannot recover 
it, annealing
of the CCDs can be attempted by heating the CCD mounting plate to 
130$^{\circ}$C. This should remove the damage, but carries a risk to the CCDs and
their connections. It will only be used as a last resort.

\subsection{Contamination}

When there is a direct path from the interior of the optical bench 
to the cold
surface of the CCD, it is possible for ice to build up on the CCD caused by
water outgassed by the spacecraft. The filter, which is at about -20$^{\circ}$C,
provides a warm primary barrier to molecules, but the gaps around the 
filter can
allow molecules to reach the CCD surface by several bounces. So far there is no
evidence of ice build-up. If this is detected later on, then the 
filter wheel can be
put in the open position and the CCDs heated to drive off the ice.


\section{Background}

The dark current and effects of light will be negligible in most circumstances
for EPIC opbservations. Cosmic rays are removed on-board by the EMCR. The
background in the image therefore consists of X-ray like events, and some
electronic pattern noise. The latter appeared in orbit, not having been seen
during ground testing, and results in a grid pattern appearing on images
integrated for a long time. It is probably generated by low level EMC but the
mechanism remains obscure. Fortunately only a few pixels are affected per
frame, and these can easily be removed from the image by a simple algorithm,
with negligible loss of X-rays. The remaining background is X-ray like and
cannot be removed. It is made up of the small residuum from the removal of 
cosmic
rays, the diffuse X-ray background, Compton interactions of cosmic and locally
generated gamma rays, and soft protons.
The soft protons can be removed or minimised by excluding periods when they
are present; this is indicated by flaring in the count-rate from the 
outer CCDs.
Non-flaring periods may represent an absence of soft
protons or simply a low and steady flux of them. The gamma ray and cosmic
ray induced background cannot be removed.
The flux appears to be twice as big as predicted for this orbit; this may be
related to the active state of the sun, or to inadequacies in the 
radiation models.
The spectra of the soft proton
and gamma induced backgrounds are much flatter than those of the diffuse X-ray 
background and of X-ray sources. So the effect on sensitivity is more
marked at the higher energies.
The cameras were designed to minimise the number of different materials that
can generate fluorescent X-ray lines. Only silicon and aluminium lines appear
in the background, at fairly low levels.


\begin{acknowledgements}
XMM-Newton is an ESA science mission with instruments and contributions 
directly funded by ESA Member States and the USA (NASA).
EPIC was developed by the EPIC Consortium led by the Principal 
Investigator, Dr. M. J. L. Turner. The consortium comprises the 
following Institutes: University of Leicester, University of 
Birmingham, (UK); CEA/Saclay, IAS Orsay, CESR Toulouse, (France); 
IAAP Tuebingen, MPE Garching,(Germany); IFC Milan, ITESRE Bologna, 
OAPA Palermo, Italy. EPIC is funded by: PPARC, CEA, CNES, DLR and
ASI. The EPIC collaboration wishes to thank 
EEV Ltd, now Marconi Advanced Technology, MMS Stevenage and LABEN 
S.p.A-Milano for their dedication to the EPIC program.

\end{acknowledgements}

\bibliography{EPIC_HW_final}

\end{document}